\documentclass{jfm}
\usepackage{graphicx}
\usepackage{epstopdf, epsfig}
\usepackage{gensymb}
\usepackage{mathrsfs}
\usepackage{amsmath}
\usepackage{layouts}

\usepackage{xcolor}
\usepackage{soul}

\shorttitle{Dynamic near-wake modulation}
\shortauthor{A. Abraham, L. A. Mart\'{i}nez-Tossas and J. Hong}

\title{Mechanisms of dynamic near-wake modulation of a utility-scale wind turbine}

\author{Aliza Abraham\aff{1,2},
  Luis A. Mart\'{i}nez-Tossas\aff{3}
 \and Jiarong Hong\aff{1,2}
 \corresp{\email{jhong@umn.edu}}}

\affiliation{\aff{1}St. Anthony Falls Laboratory, University of Minnesota,
Minneapolis, MN 55414, USA
\aff{2}Department of Mechanical Engineering, University of
Minnesota, 
Minneapolis, MN 55455, USA
\aff{3}National Renewable Energy Laboratory, Golden, CO 80401, USA}

\begin{document}

\maketitle

\begin{abstract}
The current study uses large eddy simulations to investigate the transient response of a utility-scale wind turbine wake to dynamic changes in atmospheric and operational conditions, \textcolor{blue}{as observed in previous field-scale measurements}. Most wind turbine wake investigations assume quasi-steady conditions, but real wind turbines operate in a highly stochastic atmosphere, and their operation (e.g., blade pitch, yaw angle) changes constantly in response. Furthermore, dynamic control strategies have been recently proposed to optimize wind farm power generation and longevity. Therefore, improved understanding of dynamic wake behaviors is essential. First, changes in blade pitch are investigated and the wake expansion response is found to display hysteresis as a result of \textcolor{blue}{flow} inertia. The timescales of the wake response to different pitch rates are quantified. Next, changes in wind direction with different timescales are explored. Under short timescales, the wake deflection is \textcolor{blue}{in the opposite direction} of that observed \textcolor{blue}{under quasi-steady conditions}. Finally, yaw changes are implemented at different rates, and the maximum inverse wake deflection and timescale are quantified, showing a clear dependence on yaw rate. To gain further physical understanding of the mechanism behind the inverse wake deflection, the streamwise vorticity in different parts of the wake is quantified. The results of this study provide guidance for the design of advanced wake flow control algorithms. The lag in wake response observed for both blade pitch and yaw changes shows that proposed dynamic control strategies \textcolor{blue}{must implement turbine operational changes with a timescale on the order of the rotor timescale or slower}. 
\end{abstract}


\section{Introduction}
Utility-scale wind turbines operate in highly stochastic atmospheric conditions and are subject to constant changes in wind speed, direction, and turbulence. \textcolor{blue}{Simplified models of these conditions can be developed, but they cannot be fully replicated} in the laboratory or in simulations. Moreover, \textcolor{blue}{turbine} operation (e.g., blade pitch, rotor speed, yaw angle) is constantly adapting to these changes, further increasing the complexity of the fluid-structure interactions they experience. Improved understanding of these dynamic interactions is essential for continued increases in turbine size and efficiency \citep{stevens2017,porte2018wind,Veers2019}. 

\citet{Dasari2019} first explored the effect of changes in blade pitch on the wake of a 2.5 MW wind turbine in the field, finding significant impact on wake expansion and blade tip vortex behaviour. Another recent field-scale study investigated the effect of dynamic conditions, including wind direction, blade pitch, and tip speed ratio (ratio of the speed of the blade tips to the incoming wind speed), on the wake of the same 2.5 MW turbine \citep{Abraham2020}. Using super-large-scale flow visualization with natural snowfall, this investigation revealed the significant impact of changes in atmospheric and operational conditions on wake deflection and expansion, termed dynamic wake modulation. Changes in blade pitch and tip speed ratio were shown to cause fluctuations in the wake expansion angle and, remarkably, dynamic changes in wind direction were found to deflect the wake in the opposite direction of that observed under steady conditions. Furthermore, these dynamic wake behaviours were shown to enhance mixing up to 20\%, accelerating wake recovery. While these findings provided many useful insights into wake behaviors in the field, the underlying mechanisms causing these behaviors were not clear due to the inherent limitations of field studies, including the limited field of view and the lack of control over wind conditions.

\textcolor{blue}{Laboratory-scale experiments allow for greater control over flow conditions, enabling direct observation of the transient wake response to dynamic conditions. \citet{Yu2017} used an actuator disk in a wind tunnel to investigate step changes in thrust coefficient on the loading and wake. They observed an overshoot or undershoot of wake velocity response, depending on the direction of the thrust coefficient change, attributed to passage of the vorticity shed from the disk edge. With a similar setup, \citet{Macri2021} studied step changes in yaw angle. They found a lag in wake deflection and thrust coefficient response on the order of the yaw maneuver duration. To investigate the effect of dynamic inflow on turbine blade loading, \citet{Schepers2007} implemented step changes in blade pitch on a 2-bladed, 10 m diameter turbine in the NASA-Ames wind tunnel. In this experiment, load overshoots were observed before a gradual approach to the new equilibrium. \citet{Berger2018} conducted a similar study to investigate the impact of fast changes in blade pitch on a scaled turbine model and its wake. They also recorded overshoot and undershoot of turbine loading in addition to a decaying wake velocity response attributed to flow inertia. Though these experimental studies successfully isolate individual parameters to determine their effect on the wake, the mechanisms causing the wake changes are still not clear due to their inability to directly observe all variables (e.g., vorticity) at all locations around the rotor. Furthermore, the scalability of wind tunnel experiments is limited by constraints on the physical scale and mechanical characteristics of model turbines and the challenge of replicating the complexity of atmospheric flow.}

Simulations have the potential to elucidate these mechanisms, as flow and turbine parameters can be easily controlled and modified, \textcolor{blue}{and all variables can be extracted from the domain}. A few simulation studies have investigated the effect of dynamic conditions on wind turbine performance and wake behavior. \citet{Leishman2002} reviewed the challenges involved in modelling the unsteady aerodynamics of wind turbines, and used dynamic inflow theory to estimate the time constant for flow development through a rotor as 1-1.5 rotor revolutions. This study also showed a qualitative picture of the wake response to a 30\degree{} step change in yaw angle using the vortex wake model, demonstrating that it took about 9 rotor revolutions for the wake to fully stabilize. \citet{Ebrahimi2018} simulated the effect of step changes in both wind speed and yaw angle on power production and blade loading, finding the timescale for stabilization to be 4-5 s. Finally, \citet{Andersen2018} used large eddy simulations (LES) to investigate the power output response to changes in wind speed and direction. They found a lag between peaks in thrust force and peaks in power that they attributed to generator inertia and the timescale of the controller. Further, they observed that the correlations between the turbine loads and the wake position were low, highlighting the highly dynamic behavior of the wake. 

In addition to gaining insight into wake behaviors occurring during normal utility-scale wind turbine operation, improved understanding of the dynamic wake is crucial for the design of recently proposed advanced wind farm control algorithms, including thrust optimization \citep{Goit2015,Munters2017,Shapiro2017,Munters2018b,Yilmaz2018}, yaw angle modification \citep{Gebraad2015,Raach2017,Raach2018,fleming2019,Howland2020}, and a combination of the two \citep{Munters2018,Kanev2018}. For example, \citet{Raach2017,Raach2018} used simulations to investigate the effectiveness of a lidar-based closed-loop control framework, and observed a period of inverse wake deflection in response to a 5\degree{} step change in yaw angle. \textcolor{blue}{\citet{Yilmaz2018} and \citet{Munters2018b} took advantage of changes in wake width induced by periodic blade pitch changes to induce vortex rings in the wake of a simulated turbine, thereby re-energizing the wake through their interactions. The wind farm optimization conducted by \citet{Munters2018} yielded similar behaviors for thrust and analogous yaw oscillations that induced wake meandering.} The current study employs LES to explore the underlying physical mechanisms behind these dynamic wake behaviors.

The current study focuses on developing the understanding of dynamic wake behaviors, both those occurring during normal utility-scale turbine operation and those that result from the implementation of adaptive control algorithms. The investigation uses LES to model changes in blade pitch, wind direction, and rotor yaw occurring at different rates. The timescales and mechanisms of the wake response are then described. Section \ref{sec:methods} describes the simulation methodology and the turbine model, Section \ref{sec:results} presents the results, and Section \ref{sec:conc} provides the discussion of the results and conclusions.

\section{Methodology}\label{sec:methods}
\subsection{Large eddy simulations}\label{sec:les}
\begin{figure}
  \centerline{\includegraphics{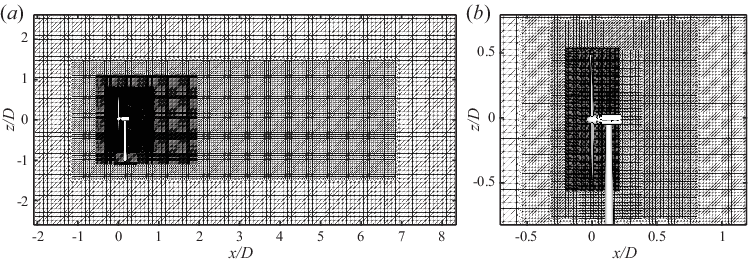}}
  \caption{(\textit{a}) The entire simulation domain, showing the grid refinement and the position of the turbine. (\textit{b}) A smaller portion of the simulation domain highlighting the refined region around the rotor. Note that the simulated turbine is represented using actuator lines. The model included here is shown for visualization purposes. \textcolor{blue}{In the current study, wake behaviors are analyzed at $x/D=0.18$ and $x/D=5$.}}
\label{fig:mesh}
\end{figure}

LES were performed using Nalu-Wind, a wind-focused fork of the incompressible flow solver developed by Sandia National Laboratories, Nalu \citep{Sprague2020}. Nalu-Wind solves the filtered Navier-Stokes equations using an unstructured \textcolor{blue}{finite-volume formulation that is second order accurate in space and time.} The deviatoric part of the resolved Reynolds stress tensor was represented using a Smagorinsky model with a Smagorinsky coefficient of $C_S=0.08$, \textcolor{blue}{based on the average of the values computed by the Lagrangian scale-dependent Smagorinsky model. \citet{Martinez-Tossas2018} shows that when using $C_S=0.08$, the turbulence characteristics (e.g., resolved Reynolds stresses) agree well with the results from the more realistic Lagrangian scale dependent model.}. \textcolor{blue}{The timestep of the simulations was chosen to be 0.025 s such that the Courant–Friedrichs–Lewy (CFL) number based on the tip velocity remained below 1.} The simulations were conducted using uniform laminar inflow with an incoming wind speed of $U_\infty=10$ m/s. Uniform inflow was selected to isolate the changes in turbine operation and their effect on the near wake. Previous simulation studies have frequently used uniform laminar inflow to separate deterministic wake behaviors from the effects of turbulent stochasticity \citep{Tsalicoglou2014,Troldborg2015,howland2016,Kim2016,Rahimi2018}. \textcolor{blue}{In addition, the near wake (the focus of the current study) is dominated by coherent structures shed from the turbine rather than random turbulence \citep{Sorensen2011}. Previous field-scale studies using high-resolution flow visualization also exhibit the dominance of coherent structures in the near wake \citep{Hong2014,Yang2016,Dasari2019,Abraham2019,Abraham2020}. Most importantly}, the results of the recent field study investigating the dynamic wake modulation phenomena explored in the current study show that these large-scale near-wake behaviors dominate above atmospheric turbulence \citep{Abraham2020}. The size of the simulation domain was 1000 m $\times$ 500 m $\times$ 500 m \textcolor{blue}{($10.4D \times 5.2D \times 5.2D$, where $D=96$ m is the turbine rotor diameter)}, and the top, bottom, and sides of the domain had symmetry boundary conditions imposed. No ground surface was modelled in order to maintain uniform inflow. The resolution farthest from the rotor was 10 m, and the mesh was refined progressively to 0.625 m directly around the rotor, producing a grid with a total of 5.6 million nodes (figure \ref{fig:mesh}). This resolution is typical of or finer than similar large-scale simulations in the literature \citep{Martinez-Tossas2018,Blaylock2019,Nielson2019}.

\subsection{Wind turbine model}
The simulated wind turbine was modelled after the University of Minnesota Eolos wind turbine, described in previous studies \citep{Hong2014,Chamorro2015,Dasari2019,Abraham2019,Abraham2020}. It is a 2.5 MW Clipper Liberty C96 three-bladed, horizontal-axis, pitch-regulated, variable speed machine with a 96 m rotor diameter ($D$) and 80 m hub height ($H_{\text{hub}}$). The turbine model was implemented in OpenFAST \citep{Sprague2020}, and the controller was tuned to match the real power curve as closely as possible. The model was then incorporated into the LES using an actuator line model, where the blades were represented by a force distribution along a line extending from the hub to each blade tip. The actuator line for each blade was discretized into 100 points, and the forces from the blades were then projected onto the flow using a Gaussian function \citep{Sorensen2002,martinez2014large}. The blade forces were calculated using lift and drag coefficient look-up tables determined by the specifications of the Eolos turbine used for the field studies. The nacelle and tower were included in the model as additional body forces, as these structural components have been shown to significantly impact the near wake \citep{Abraham2019}. \textcolor{blue}{The tower was modeled as a drag force discretized into 50 points to allow for variation in cross-sectional diameter along the height. As with the blades, the forces were projected onto the flow using a Gaussian function. The nacelle was modeled as a Gaussian drag body force with a frontal area of $A_n=20$ m\textsuperscript{2} and a drag coefficient of $C_{D,n}=0.5$. The width of the Gaussian kernel was determined as a function of $A_n$ and $C_{D,n}$ to ensure the appropriate momentum thickness of the generated wake \citep{MartinezTossas2017}: $\epsilon_D=\sqrt{2C_{D,n}A_n/\pi}$.} Previous studies have shown that the actuator line model can successfully reproduce the flow features that dictate the large-scale near-wake behaviours investigated in the current study, i.e., near-wake flow fields and coherent structures \citep{Sanderse2011,Lu2011,Kang2014,martinez2014large,Nilsson2015}. \textcolor{blue}{Tip loss corrections have not been included in the current study. \citet{Sarlak2016} showed that the wake of a turbine modeled using actuator lines is not significantly modified by the use of tip loss corrections for low to intermediate tip speed ratios, such as those used in the current study.}

\subsection{Wake detection and fitting}\label{sec:fit}
\begin{figure}
  \centerline{\includegraphics{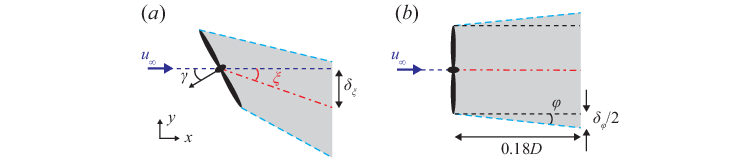}}
  \caption{ (\textit{a}) Schematic of the turbine wake showing the definition of the yaw misalignment angle, $\gamma$, the wake deflection angle, $\xi$, and the wake deflection magnitude, $\delta_{\xi}$, from the top (positive-$z$) view. Coordinate axes are also shown. (\textit{b}) Schematic of the turbine wake showing the definition of the wake expansion angle, $\varphi$, and the wake expansion magnitude, $\delta_{\varphi}$. The incoming wind is indicated by $u_\infty$ and the measurement plane is located $0.18D$ downstream of the rotor.}
\label{fig:def}
\end{figure}

The current study focuses on the near wake \citep[$0.18D$ downstream, selected to enable comparison with field-scale experimental results from][]{Abraham2020} where the turbine-induced flow modulation originates. To quantify the centerline and width of the wake, the velocity at the cross-section is fit with a two-dimensional Gaussian function,
\begin{equation}
  u(y,z)=A*\exp \left( -\frac{(y-y_c)^2+(z-z_c)^2}{2\sigma^2} \right) +U_\infty,
\end{equation}
where $(y_c,z_c)$ is the center of the Gaussian, $\sigma^2$ is the variance, and $A$ is a fitting coefficient \citep{BASTANKHAH2014,Quon_2020}. The Gaussian wake model is used because it minimizes the number of fitting coefficients, whereas other wake models, particularly those designed for the near wake, require many additional fitting parameters \citep[e.g.,][]{Keane2016,Ishihara2018,Cathelain2020}. The center of the wake cross-section is taken to be the center of the Gaussian fit, which is used to determine the wake deflection angle, $\xi$ (figure \ref{fig:def}\textit{a}). Often the far wake width is defined as the 95\% confidence interval of the Gaussian velocity deficit profile \citep{Aitken2014,Doubrawa2017}. However, the near wake velocity deficit has steeper edges than in the far wake, so the 95\% confidence interval would severely overestimate the near wake width. By comparing the fitted Gaussian to the near-wake velocity deficit obtained from the LES, we found that the 91\% confidence interval provides a more accurate definition of the near wake width. The justification of this selection is described in more detail in Appendix \ref{appA}. The wake width is then converted to the wake expansion angle, shown in figure \ref{fig:def}(\textit{b}), using the following relation:
\begin{equation}
  \varphi=\arctan \left( \frac{1.7 \sigma -\frac{D}{2}}{0.18D} \right),
\end{equation}
where the factor of 1.7 corresponds to the 91\% confidence interval and $0.18D$ is the downstream distance of the measurement plane.

\section{Results}\label{sec:results}
We now present results for the turbine wake response to three types of operational changes: 1) variations in expansion caused by blade pitch, 2) redirection from wind direction changes, and 3) wake deflection due to rotor yaw. Sample images of the wake velocity during each of these scenarios are shown in figure \ref{fig:stills}. 

\begin{figure}
  \centerline{\includegraphics{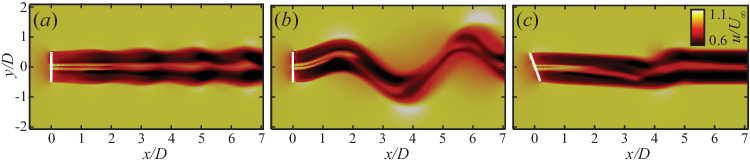}}
  \caption{Sample timesteps of the wake streamwise velocity at hub height from the top (positive-$z$) view during the simulations modeling changes in (\textit{a}) blade pitch, (\textit{b}) wind direction, and (\textit{c}) rotor yaw. \textcolor{blue}{White lines indicate the position and orientation of the rotor.}}
\label{fig:stills}
\end{figure}

\subsection{Blade pitch}
\begin{figure}
  \centerline{\includegraphics{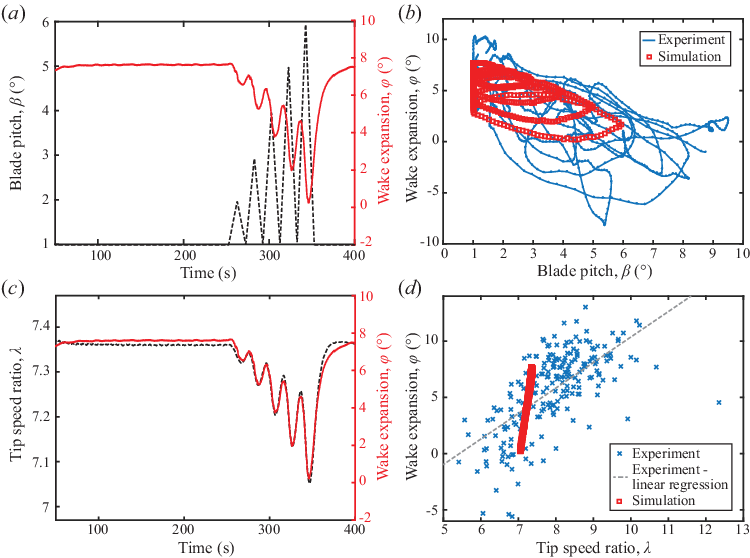}}
  \caption{(\textit{a}) Prescribed changes in blade pitch (dashed black line) and the resulting changes in wake expansion (solid red line). (\textit{b}) Instantaneous wake expansion angle versus blade pitch from  the simulation (red squares) and a sample sequence of 800 s from the experiment from \citet{Abraham2020} (blue line). (\textit{c}) Time series of tip speed ratio changes (dashed black line) caused by changes in blade pitch and resulting wake expansion (solid red line). (\textit{d}) Instantaneous wake expansion angle versus tip speed ratio from the simulation (red squares) and from periods of the experimental dataset from \citet{Abraham2020} where the wind is above the rated speed and the blade pitch is changing (blue crosses). A linear regression fit to the experimental data is included to facilitate comparison with the simulation data.}
\label{fig:pitch}
\end{figure}

First the effect of changes in blade pitch ($\beta$) on the near wake is investigated. Most utility-scale wind turbine controllers increase the pitch of the blades when the incoming wind is above the rated wind speed to reduce the angle of attack of the airfoils and decrease the thrust force of the rotor as a way to regulate the structural loading. To examine the effect of this process on the wake, a time series of pitch changes at different rates, \textcolor{blue}{similar to those observed for a utility-scale turbine,} is implemented in the OpenFAST turbine controller. The pitch first increases at each rate for 10 s, then decreases at the same rate for 10 s, as shown in figure \ref{fig:pitch}(\textit{a}) along with the resulting changes in wake expansion angle. The wake expansion varies inversely with changes in blade pitch due to the aforementioned reduction in thrust caused by lowering the angle of attack. The same relationship is observed in the field experiment from \citet{Abraham2020}. Interestingly, the correlation is not linear, rather some hysteresis occurs in the wake response. This hysteresis can also be observed in the field data, as evidenced by the fact that there are multiple different values of wake expansion for a given blade pitch angle during a sample sequence of 800 s (figure \ref{fig:pitch}\textit{b}). The same characteristic looping behavior is observed in the relationship between blade pitch and wake expansion from the field data as that observed in the simulation results, though the loops are not as regular as those from the simulations due to the many different factors influencing wake behaviour in the field. Because individual variables are difficult to isolate in the field, the cause of this observation could not be confirmed with certainty. In the simulation, the cause of such hysteretic response becomes clear when comparing the wake expansion with the tip speed ratio, $\lambda=\frac{\omega D}{2u_\infty}$ where $\omega$ is the angular velocity of the rotor, which changes in response to changes in blade pitch. Figure \ref{fig:pitch}(\textit{c}) shows the strong correlation between $\lambda$ and $\varphi$, which suggests that the wake is actually responding to the change in rotor speed rather than the change in blade pitch itself. A similar trend is observed in the experimental data, conditionally sampled for periods where the wind is above the rated speed and the blade pitch is changing, though with a wider range of $\lambda$ due to larger changes in $u_\infty$ and much more variability due to the dynamic atmospheric  wind speed and direction (figure \ref{fig:pitch}\textit{d}). \textcolor{blue}{Note that the results of the simulation are repeatable because of the use of uniform inflow, while the experimental data is not deterministic due to the multitude of uncontrolled stochastic variables in the field. Still, the trends observed in the experimental data are robust, as described in more detail in \citet{Abraham2020}.} The relationship between $\lambda$ and $\varphi$ is stronger than that between $\beta$ and $\varphi$ because the reduction in \textcolor{blue}{lift force on the blades} due to the increase in pitch causes the rotor to slow down, which causes the wake expansion to decrease. This reduction in rotor speed lags slightly behind the increase in blade pitch due to the inertia of the \textcolor{blue}{flow. \citet{Schepers2007} described the effect of flow inertia to explain the observed lag in turbine response to a step increase in blade pitch. He attributed the lag to the time it takes to accelerate the flow after a decrease in axial induction}. \citet{Ebrahimi2018} observed a similar delay in thrust change after a step change in wind speed. A linear regression fit is applied to the relationship between $\lambda$ and $\varphi$ from the experiment, revealing a shallower slope than that of the simulation data (figure \ref{fig:pitch}\textit{d}). This difference in slope is attributed to the smoothing applied to the experimental wake boundary data to account for turbulent fluctuations, which tends to obscure more extreme values. The experimental data also includes many additional sources of uncertainty, including concurrent changes in multiple parameters that may interact, and the measurement of incoming velocity at the turbine nacelle, which is within the induction zone of the turbine \citep{Li2020}. However, it is remarkable that the overall trends from the experiment and simulations still match well in spite of these discrepancies, indicating that the large-scale wake behaviours dominate over the effects of smaller-scale turbulence.

\begin{figure}
  \centerline{\includegraphics{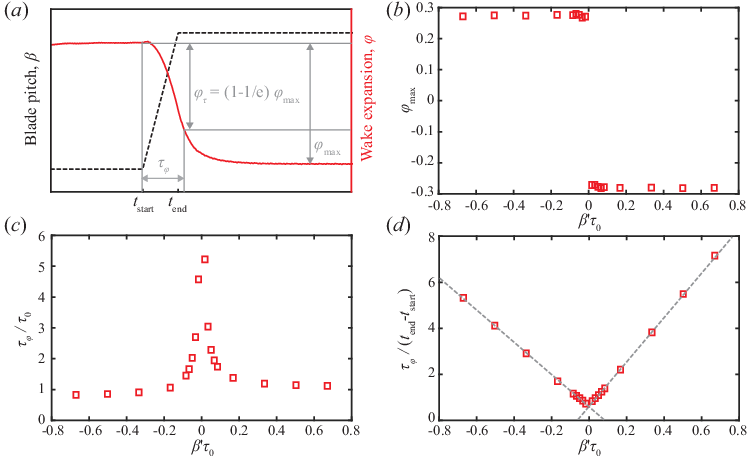}}
  \caption{{(\textit{a}) Sample change in blade pitch angle at a constant rate (dashed black line) and the resulting wake expansion response (solid red line). The wake expansion timescale, $\tau_{\varphi}$, and maximum change in wake expansion, $\varphi_{\text{max}}$, are defined in grey. (\textit{b}) Relationship between \textcolor{blue}{normalized} blade pitch rate and maximum wake expansion change. (\textit{c}) Relationship between \textcolor{blue}{normalized} pitch rate and \textcolor{blue}{normalized} wake expansion timescale. (\textit{d}) Relationship between \textcolor{blue}{normalized} pitch rate and \textcolor{blue}{wake expansion timescale normalized by the duration of the blade pitch maneuver. The grey dashed lines show linear regression fits to the data for $\beta'\tau_0 < 0$ and $\beta'\tau_0 > 0$.}}}
\label{fig:pitch_rate}
\end{figure}

To quantify the timescale of the wake expansion response, changes in blade pitch angle at different rates \textcolor{blue}{($\beta'$)} are simulated. Blade pitch is varied from $1\degree$ (the default pitch angle of the Eolos turbine when operating below the rated wind speed) to $7\degree$ at several different rates between 0.1 \degree{}/s and 4 \degree{}/s, and from $7\degree$ to $1\degree$ at rates between -0.1 \degree{}/s and -4 \degree{}/s. \textcolor{blue}{Such step changes are used to replicate the blade pitching behavior of an operational turbine, which cannot change the blade pitch instantaneously.} Figure \ref{fig:pitch_rate}(\textit{a}) shows a sample time sequence of the change in wake expansion in response to an increase of blade pitch at a constant rate. The wake begins expanding when the pitch begins changing, then asymptotically approaches its maximum change in expansion, $\varphi_{\text{max}}$. \textcolor{blue}{To facilitate physical interpretation of the results, values are normalized using the timescale $\tau_0=D/U_\infty$ and angles are converted from degrees to radians.} Figure \ref{fig:pitch_rate}(\textit{b}) shows that $\varphi_{\text{max}}$ is independent of pitch rate, and that with a $6\degree$ \textcolor{blue}{(0.10 radian)} change in blade pitch, the wake expansion always changes by $\sim16\degree$ \textcolor{blue}{(0.28 radians)}. On the other hand, the duration of the wake expansion change is strongly dependent on pitch rate. To quantify this effect, a wake expansion timescale ($\tau_{\varphi}$) is defined as in exponential decay, i.e., as the time for the change in wake expansion to reach $\varphi_{\tau}=(1-1/e)\varphi_{\text{max}}$. \textcolor{blue}{\citet{Berger2018} similarly observed exponential decay of wake velocity in response to a step change in rotor thrust.} Figure \ref{fig:pitch_rate}(\textit{c}) shows that $\tau_{\varphi}$ is significantly higher at slow pitch rates \textcolor{blue}{($\tau_{\varphi}/\tau_0\sim 5$)} compared to fast pitch rates \textcolor{blue}{($\tau_{\varphi}/\tau_0\sim 1$)}. \textcolor{blue}{Figure \ref{fig:pitch_rate}(\textit{d}) further elucidates this trend by scaling the wake expansion timescale by the pitch maneuver duration ($t_\text{end}-t_\text{start}$). This scaling reveals that $\tau_{\varphi}/(t_\text{end}-t_\text{start})$ is linearly dependent on $\beta'\tau_0$, highlighting that as the pitch rate increases, the lag in wake response increases, indicating an increased deviation from quasi-steady behavior. Some asymmetry is observed between increasing and decreasing blade pitch, where the timescales tend to be larger for increasing pitch. This trend is consistent with the explanation provided by \citet{Schepers2007} for the difference between an upward and downward pitching step: For an increase in blade pitch, the axial induction is initially large, leading to lower wake velocity and advection time than for a decrease in blade pitch, where the wake velocity is initially higher.} To provide a sense of the typical blade pitch rate of a utility-scale wind turbine operating in the field, the mean pitch rate implemented by the Eolos turbine during the experiment described in \citet{Abraham2020} was 0.4 \degree{}/s \textcolor{blue}{($\beta'\tau_0=0.04$)}, with a maximum and minimum rate of 2.4 \degree{}/s \textcolor{blue}{($\beta'\tau_0=0.26$)} and 0.2 \degree{}/s \textcolor{blue}{($\beta'\tau_0=0.02$)}, respectively. These findings have important implications for the design of advanced wind farm control algorithms that use blade pitch to regulate turbine axial induction \citep[e.g.,][]{Yilmaz2018}, which cannot expect the wake to react immediately to changes in blade pitch.

\subsection{Wind direction}
\begin{figure}
  \centerline{\includegraphics{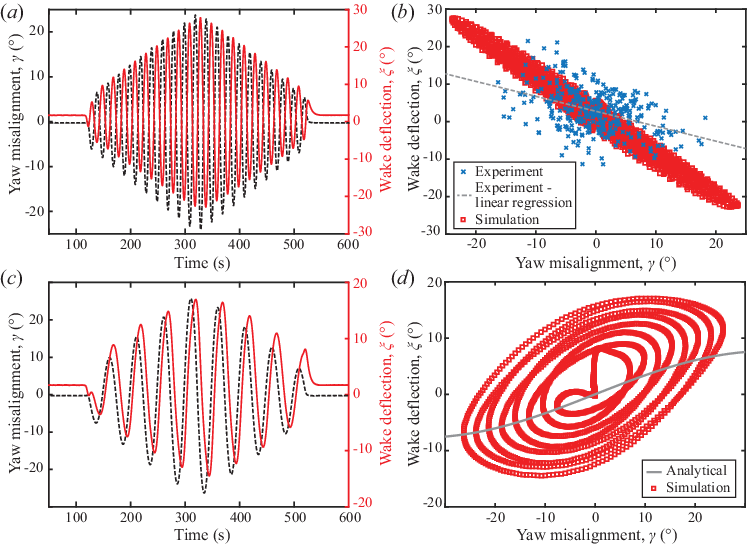}}
  \caption{(\textit{a}) Prescribed changes in wind direction with a 20 s period (dashed black line) and the resulting changes in spanwise wake deflection (solid red line). (\textit{b}) Instantaneous wake deflection angle versus wind direction for the experiment from \citet{Abraham2020} (blue crosses) and the simulation (red squares). A linear regression fit to the experimental data is included to facilitate comparison with the simulation data. (\textit{c}) Prescribed changes in wind direction with a 50 s period (dashed black line) and the resulting changes in spanwise wake deflection (solid red line). (\textit{d}) Analytical solution for the steady wake deflection angle from \citet{Jimenez2010} (grey line) compared to the instantaneous wake deflection versus wind direction from the simulation (red squares).}
\label{fig:dir}
\end{figure}

In \citet{Abraham2020}, a clear correlation was observed between instantaneous incoming wind direction and spanwise wake deflection, though it was in the opposite direction of that described in several previous studies conducted under steady conditions \citep[e.g.,][]{Jimenez2010,Bastankhah2016,Shapiro2018}. Here the mechanism behind such conflicting results is investigated in more detail. First, the spanwise component of the wind is varied sinusoidally with a 20 s period while the rotor direction is held constant, providing the turbine with dynamic yaw misalignment ($\gamma$) similar to the conditions experienced in the field due to the stochasticity of the atmospheric flow (figure \ref{fig:dir}\textit{a} and supplementary movie 1). The resulting instantaneous wake deflection is shown to follow the trend observed in the field (figure \ref{fig:dir}\textit{b}). Note that, as in the wake expansion case, the slope of the wake deflection response to yaw misalignment angle is shallower for the experimental data than the simulation data. Once again this discrepancy is attributed to the turbulence and uncertainties of the experimental data. Additionally, as will be described in more detail in the following sentences, the wake deflection response is dependent on the timescale of wind direction changes. Atmospheric wind experienced by turbines in the field fluctuates across a range of timescales, whereas the simulated wind direction changes with a constant period. 

The period of wind direction changes is increased to 50 s, and a different trend is observed compared with the 20 s period (figure \ref{fig:dir}\textit{c} and supplementary movie 2). With these slower wind direction changes, the wake deflection more closely follows the analytical relationship defined in \citet{Jimenez2010} for steady yaw error, though with some hysteresis (figure \ref{fig:dir}\textit{d}). This hysteresis is caused by a lag in the wake response to changes in yaw error. These results suggest the existence of a characteristic wake response timescale during which the wake deflection transitions from the transient opposite deflection response to the steady analytical response. \textcolor{blue}{The dependencies of this timescale will be explored in the following section using step changes in $\gamma$ at different rates.} Note that the wake deflection response is asymmetrical, with a larger deflection on the positive side than the negative side, likely caused by wake rotation \citep{Bastankhah2016}. The turbine modelled in the current study rotates in the counterclockwise direction. For a clockwise rotating turbine, the negative deflection may be larger than the positive deflection.

\subsection{Rotor yaw}
\begin{figure}
  \centerline{\includegraphics{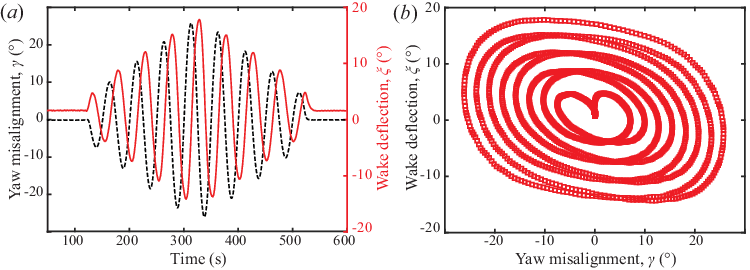}}
  \caption{(\textit{a}) Prescribed sinusoidal changes in rotor yaw angle with a period of 50 s, matching the changes in wind direction shown in figure \ref{fig:dir}(\textit{c}), (dashed black line) and the resulting changes in spanwise wake deflection (solid red line). (\textit{b}) Instantaneous wake deflection angle versus yaw misalignment angle.}
\label{fig:yaw}
\end{figure}

Using changing wind direction to investigate yaw error includes the coupling of two effects: yaw misalignment induced wake deflection and changes in the direction of wake advection by the surrounding flow. Though this coupling is representative of the wake response in the field under stochastic atmospheric flow, these effects must be separated to gain further insight into the flow modulation caused by the turbine. Therefore, the same yaw misalignment sequence as that shown in figure \ref{fig:dir}(\textit{c}) is implemented by changing the rotor yaw angle while keeping the wind direction constant (figure \ref{fig:yaw}\textit{a}). In this scenario, the lag in wake response is longer than previously observed for the changing wind direction scenario with the same period (50 s). This discrepancy is attributed to the fact that, when the wind direction changes, the surrounding flow advects the wake in the spanwise direction, pushing it towards the steady yaw response. \textcolor{blue}{Additionally, the streamwise velocity must be kept constant to ensure constant mass flow through the simulation domain, so spanwise velocity is varied in the wind direction simulation. This increase in spanwise velocity causes the wind velocity magnitude to increase. \citet{Jimenez2010} showed that the spanwise force exerted by the turbine on the wake is dependent on the square of the inflow velocity magnitude under steady conditions. Because the velocity magnitude is larger in the wind direction simulation during periods of yaw misalignment than in the rotor yaw simulation, the force pushing the wake towards steady-state deflection is stronger in the wind direction simulation. Therefore, the dynamic wake deflection approaches steady-state behavior more quickly than in the rotor yaw simulation.} Though the timescales of the wake response in the two cases are different, the general trends observed are consistent, including the magnitude of the inverse wake deflection. \textcolor{blue}{To isolate the effect of dynamic wake modulation from that of wake advection by ambient wind, and to facilitate the extension of these findings to wake control applications, the following analysis uses rotor yaw angle changes to investigate dynamic spanwise wake deflection in more detail.}

\begin{figure}
  \centerline{\includegraphics{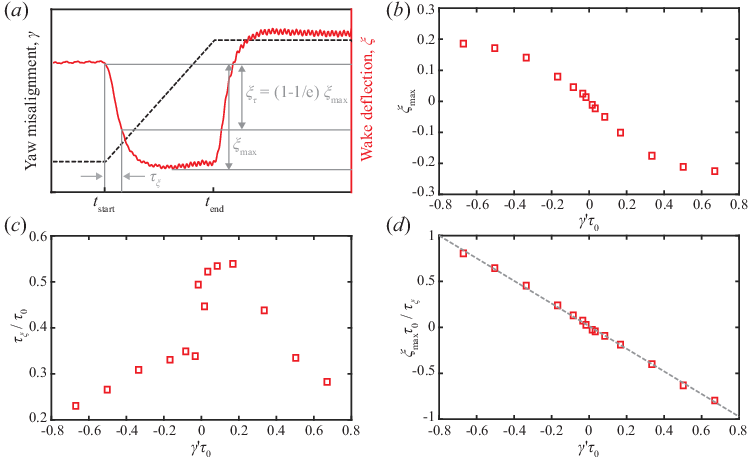}}
  \caption{(\textit{a}) Sample change in yaw angle at a constant rate (dashed black line) and the resulting wake deflection response (solid red line). The wake deflection timescale, $\tau_{\xi}$, and maximum inverse wake deflection, $\xi_{\text{max}}$, are defined in grey. (\textit{b}) Relationship between \textcolor{blue}{normalized} yaw rate and maximum inverse wake deflection. (\textit{c}) Relationship between \textcolor{blue}{normalized} yaw rate and \textcolor{blue}{normalized} inverse wake deflection timescale. (\textit{d}) Relationship between yaw rate and \textcolor{blue}{$\xi_{\text{max}}\tau_0/\tau_{\xi}$}. The dashed grey line shows the linear regression fit to the data.}
\label{fig:time}
\end{figure}

To quantify the inverse wake deflection magnitude and timescale more precisely, changes in rotor direction with constant yaw rates are investigated. The yaw angle is varied from $0\degree$ to $20\degree$ at several different rates \textcolor{blue}{($\gamma'$)} between 0.1 \degree{}/s and 4 \degree{}/s, and from $20\degree$ to $0\degree$ at rates between -0.1 \degree{}/s and -4 \degree{}/s. \textcolor{blue}{As with blade pitch, step changes are used to replicate the yawing behavior of an operational turbine, which cannot change the rotor angle instantaneously.} Figure \ref{fig:time}(\textit{a}) shows a sample time sequence of wake deflection with a constant yaw rate. Under this condition, the wake first deflects in the direction opposite of the yaw change, then deflects in the same direction as the yaw change for the remaining duration of the rotor motion, consistent with the results of previous studies \citep{Leishman2002,Raach2017,Raach2018}. \textcolor{blue}{Results are once again normalized by $\tau_0$ and angles are converted to radians.} Figure \ref{fig:time}(\textit{b}) quantifies the magnitude of the maximum inverse wake deflection, $\xi_{\text{max}}$, showing a clear dependence on yaw rate. This deflection is not insignificant; at the largest yaw rates, it reaches 75\% of the rotor misalignment angle. However, the duration of this deflection decreases with increasing yaw rate magnitude. The inverse wake deflection timescale ($\tau_\xi$) is defined in the same way as the wake expansion timescale, i.e., as the time for the wake deflection to reach $\xi_{\tau}=(1-1/e)\xi_{\text{max}}$. \textcolor{blue}{The value of $\tau_\xi/\tau_0$ generally increases with decreasing yaw rate magnitude}, though the peak occurs at a \textcolor{blue}{normalized} yaw rate \textcolor{blue}{$\gamma'\tau_0 > 0$}. This asymmetry is also likely caused by wake rotation, and may be reversed in the case of a clockwise rotating turbine. Note that $\tau_\xi$ ranges from \textcolor{blue}{$0.23\tau_0$ to $0.54\tau_0$} for the yaw rates investigated here, and a yaw rate of 0.5 \degree{}/s (\textcolor{blue}{$\gamma'\tau_0=0.08$,} typical of e.g., the Eolos turbine) results in \textcolor{blue}{$\tau_\xi=0.53\tau_0$} when the yaw is increasing and \textcolor{blue}{$\tau_\xi=0.35\tau_0$} when it is decreasing. These values show that the wake takes a significant amount of time to stabilize in response to changes in yaw angle, which has important implications for the design of yaw-based wind farm optimization algorithms. This crucial point will be discussed further in Section \ref{sec:conc}. When an inverse wake deflection rate is defined as \textcolor{blue}{$\xi_{\text{max}}\tau_0/\tau_\xi$}, another interesting trend is revealed. This quantity is linearly dependent on \textcolor{blue}{$\gamma'\tau_0$}, with a nearly one-to-one relationship (the linear regression slope is -1.2 and intersects the ordinate axis at \textcolor{blue}{$\xi_{\text{max}}\tau_0/\tau_\xi=0.01$}). This relationship clearly demonstrates the tradeoff between inverse wake deflection magnitude and duration under different yaw rates. \textcolor{blue}{The physical explanation for this relationship will be further elucidated in Section \ref{sec:vort}. Note that these yaw changes are implemented under turbine operational conditions that are otherwise optimized by the turbine controller (i.e., blade pitch is set to the value determined to maximize power based on the wind speed, as it would be for an operational utility-scale turbine). The effect of yaw changes under sub-optimal operating conditions has not been investigated. However, we expect a reduction in thrust coefficient to increase the inverse wake deflection timescale, as \citet{Jimenez2010} showed that the steady wake deflection response weakens with reduced thrust coefficient.}

\subsection{Vorticity analysis during dynamic yaw misalignment} \label{sec:vort}
Further insight into the mechanism behind these wake behaviors can be gained by looking at the vorticity in the wake. Previous studies have shown that streamwise vorticity plays an important role in the behavior of the yawed wake, leading to wake deflection and the curled wake shape that develops downstream \citep{howland2016,Bastankhah2016,Shapiro2018,Martinez-Tossas2019,Zong2020}. When streamwise vortices with opposing signs form on the top and bottom halves of the wake, they induce a spanwise velocity. Based on these studies, the following analysis focuses on the streamwise component of the vorticity vector. In the current study, after the turbine yaws, negative streamwise vorticity ($\omega_x$) is observed in the top half of the wake and positive streamwise vorticity is observed in the bottom half (see supplementary movie 3 \textcolor{blue}{\textemdash{} stills in Appendix \ref{appB}}), consistent with the steady yawed wake behavior described in \citet{Zong2020}. More interestingly, during the positive yaw maneuver, a short period of increased vorticity is observed in the top-left quadrant (I) of the wake and a period of decreased vorticity is observed in the bottom-right quadrant (IV) of the wake (figure \ref{fig:vort}\textit{a} and supplementary movie 3 \textcolor{blue}{\textemdash{} stills in Appendix \ref{appB}}). The maximum streamwise vorticity in sections of each of the four quadrants of the wake located at $z=\pm 0.3D$, $0.2D < y < 0.5D$ or $-0.5D < y < -0.2D$, and $0.5D < x < 0.6D$ compared to the theoretical vorticity trend excluding the transient effect is plotted over time in figure \ref{fig:vort}(\textit{b}), demonstrating the observed vorticity changes. Without the transient effect, the streamwise vorticity would be expected to transition smoothly from its initial value to the reduced value in the top half of the wake (quadrants I and II) and the increased value in the bottom half (quadrants III and IV). However, the transient effect shifts the vorticity in the opposite directions in quadrants I and IV. This transient vorticity shift can be explained by the angle of the vortices shed from the turbine blade tips during the yaw maneuver, where one side of the rotor is moving upstream and the other is moving downstream. As explained in \citet{Zong2020}, the angle of the blade tip vortex trajectory determines the streamwise component of the vorticity, which dictates the deflection direction of the wake. When the rotor is turning, the side moving upstream (the left side in figure \ref{fig:vort}) is experiencing a larger streamwise velocity relative to the flow, increasing the streamwise component of the vorticity on that side (supplementary movie 3). On the side moving downstream, the relative velocity is reduced, along with the streamwise vorticity component. As the yaw angle increases, the steady yaw angle effect described in \citet{Zong2020} dominates over the transient effect, leading to the expected steady-state vorticity distribution. In quadrants I and IV the transient effect is in the opposite direction of the steady effect, while the two effects are in the same direction in quadrants II and III. Therefore quadrants I and IV contribute to the opposite wake deflection while II and III do not. When the yawing maneuver ends and the two sides of the rotor experience the same relative streamwise velocity, the transient effect disappears altogether. 

\begin{figure}
  \centerline{\includegraphics{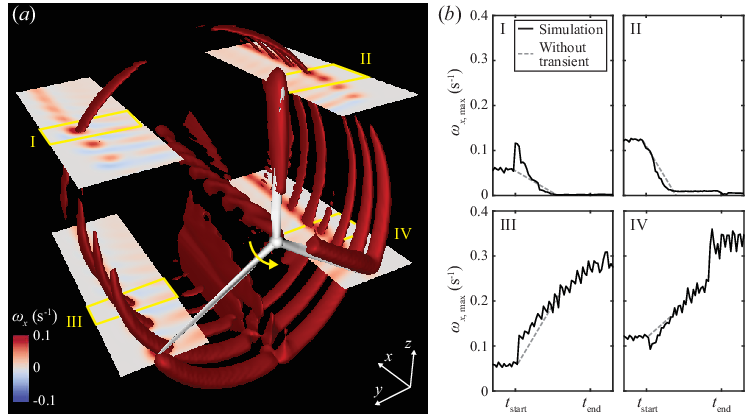}}
  \caption{(\textit{a}) Isocontour of streamwise vorticity with cut planes in each quadrant. The isocontour value is $\omega_x = 0.1$ s\textsuperscript{-1}. Note that the turbine tower and nacelle are included in the simulated turbine model, though only the rotor is shown here for clarity. The yellow arrow near the hub indicates the direction of the yaw. The yellow boxes labeled with roman numerals indicate the regions plotted in (\textit{b}), which shows the maximum streamwise vorticity component within each region over time compared to the theoretical vorticity trend excluding the transient effect. The plotted regions are located at $z=\pm 0.3D$, $0.2D<y<0.5D$ or $-0.5D<y<-0.2D$, and $0.5D<x<0.6D$.}
\label{fig:vort}
\end{figure}

\begin{figure}
  \centerline{\includegraphics{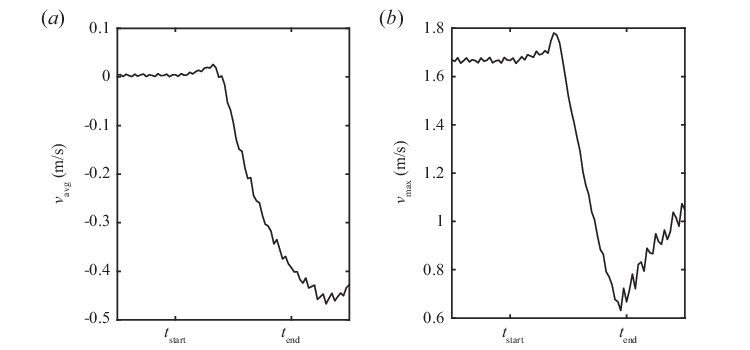}}
  \caption{(\textit{a}) Spatially averaged and (\textit{b}), maximum spanwise velocity ($v$) over the rotor area in the \textit{y-z} plane $0.5D$ downstream of the turbine during a yawing maneuver.}
\label{fig:span}
\end{figure}

Though the value of the streamwise vorticity remains positive in all four quadrants during the transient wake response, the difference in vorticity between the top and bottom of the wake is enough to induce a positive spanwise velocity and corresponding inverse wake deflection. Both the spatially averaged spanwise velocity ($v_{\mathrm{avg}}$, figure \ref{fig:span}\textit{a}) and the maximum spanwise velocity ($v_{\mathrm{max}}$, figure \ref{fig:span}\textit{b}) within the rotor area at a \textit{y-z} plane located $0.5D$ downstream show a peak just after the yawing maneuver begins, indicating a transient inverse wake deflection response. Subsequently, the spanwise velocity decreases due to the steady yaw response. As the streamwise vorticity is responsible for the negative spanwise velocity and wake deflection during the steady yaw response \citep{Zong2020}, so our results indicate the opposite streamwise vorticity shift shown in figure \ref{fig:vort} causes the positive spanwise velocity and wake deflection that are characteristic of the transient inverse wake deflection response. Note that the magnitude of the transient spanwise velocity peak increases with downstream distance, as does the inverse vorticity shift. The values at $x=0.5D$ are presented to highlight the velocity and vorticity trends that develop directly after the flow passes through the rotor.

\begin{figure}
  \centerline{\includegraphics{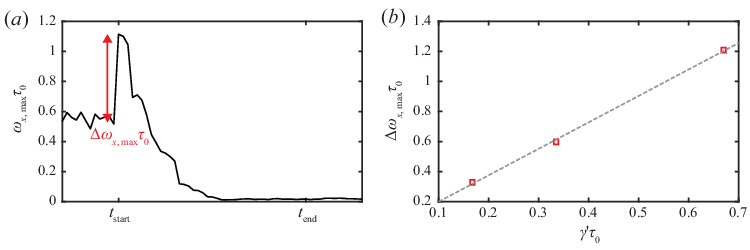}}
  \caption{\textcolor{blue}{Transient vorticity shift dependence on yaw rate, inluding (\textit{a}) the definition of the vorticity shift magnitude, $\Delta\omega_{x,\text{max}}\tau_0$, and (\textit{b}) a plot of $\Delta\omega_{x,\text{max}}\tau_0$ versus normalized yaw rate, $\gamma'\tau_0$. The dashed gray trendline shows the linear regression of the data in (\textit{b}).}}
\label{fig:vort_rate}
\end{figure}

\textcolor{blue}{The magnitude of the transient vorticity shift is directly dependent on yaw rate, as shown in figure \ref{fig:vort_rate}. This linear relationship sheds light on the clear dependence of the inverse wake deflection rate, $\xi_\text{max}\tau_0/\tau_\xi$, on yaw rate shown in figure \ref{fig:time}. The inverse wake deflection timescale $\tau_\xi$ depends on yaw rate because higher yaw rates execute a yawing maneuver of the same magnitude more quickly, but the magnitude of the inverse wake deflection $\xi_\text{max}$ is also affected by yaw rate because the strength of the vorticity generated at the rotor is determined by yaw rate. This trend is in contrast to that observed for blade pitch in figure \ref{fig:pitch_rate}, because the magnitude of the maximum wake expansion change $\varphi_\text{max}$ is unaffected by pitch rate. The transient wake expansion response to dynamic blade pitch changes is simply a slow approach to steady-state behavior, whereas the wake deflection response to yaw changes occurs in the opposite direction of the steady-state behavior.}

The vertical component of vorticity also changes when the rotor is yawing, with the left side producing stronger vorticity than the right side, as with the streamwise component. This imbalance in vertical vorticity may also induce a positive spanwise velocity, contributing to the observed spanwise wake deflection, though it would likely be much smaller than that generated by the streamwise component of vorticity \citep{Leweke2016}. Previous field-scale flow visualization studies show strong, clear tip vortices within the near-wake of utility-scale wind turbines \citep{Hong2014,Yang2016,Dasari2019,Abraham2019,Abraham2020}, so turbulence and shear are not expected to significantly \textcolor{blue}{inhibit} the aforementioned mechanism in the field. Indeed, the same inverse wake deflection described above was observed in the field data from \citet{Abraham2020}. \textcolor{blue}{That being said, figure \ref{fig:dir} shows that the magnitude of the resulting wake deflection is damped by the increased mixing caused by these atmospheric effects.}

\textcolor{blue}{Based on the analysis of the spanwise wake deflection, one might expect a similar trend in the vertical direction due to the differences in streamwise vorticity between the left and right sides of the wake. However, no significant transient vertical wake shift is observed during the yawing maneuver. This lack of vertical deflection is attributed to the large number of competing effects acting on the wake in the vertical direction. First, the interaction between wake rotation and the ground has been shown to substantially influence the vertical displacement of the wake center during yaw \citep{Bastankhah2016}. Furthermore, \citet{Kleusberg2020} showed that the angle of attack of the blades varies between the top and bottom of the rotor when the turbine is yawed, introducing an additional source of vertical asymmetry. Finally, differences in spanwise vorticity between the top and bottom of the rotor may contribute to changes in the vertical velocity field of the wake. The combination of these effects, along with the spanwise asymmetry in streamwise vorticity observed in figure \ref{fig:vort}, balance such that no significant transient vertical wake deflection is observed.}

\subsection{Downstream propagation}\label{sec:down}
\begin{figure}
  \centerline{\includegraphics{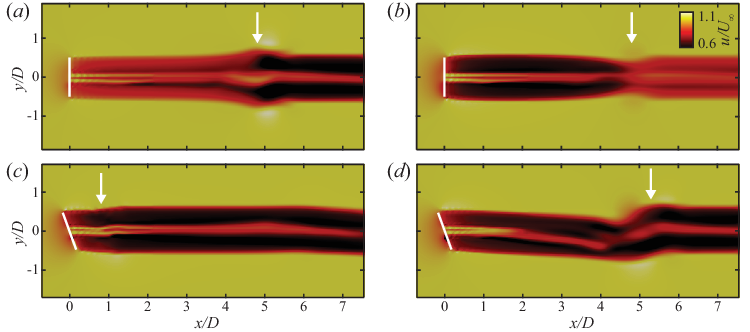}}
  \caption{Sample timesteps of the wake streamwise velocity at hub height from the top (positive-$z$) view 50 s after (\textit{a}) an increase and (\textit{b}) a decrease in blade pitch. The resulting changes in wake expansion have advected several times the rotor diameter downstream, indicated by the white arrows. Sample timesteps of the same quantity (\textit{c}) 10 s and (\textit{d}) 60 s after a change in rotor yaw. \textcolor{blue}{The position and orientation of the turbine rotor is marked by white lines and} the location of the resulting inverse wake deflection at each timestep is indicated by white arrows.}
\label{fig:down}
\end{figure}

In order to apply the findings of the current study to the development of control strategies, the extension of these near-wake timescales to the far wake must be investigated. Qualitatively, it is clear that these dynamic wake modulation behaviours persist downstream. In the wake expansion case, an interesting phenomenon is observed as the wake advects downstream after a change in blade pitch. When the blade pitch increases, the velocity deficit decreases due to a reduction in lift on the blades, so the upstream part of the wake moves faster than the downstream part, leading to a bulge at the point of blade pitch change (figure \ref{fig:down}\textit{a}). When the blade pitch decreases, the opposite occurs and the wake pinches off (figure \ref{fig:down}\textit{b}). \textcolor{blue}{These same wake width changes have been exploited to enhance wake mixing in previous studies investigating dynamic induction control of wind farms \citep{Munters2018,Munters2018b,Yilmaz2018,Frederik2020}.} When the rotor yaw changes, the disturbance due to inverse wake deflection is also observed to persist and strengthen as it moves into the far wake (figure \ref{fig:down}\textit{c} and \textit{d}).

\begin{figure}
  \centerline{\includegraphics{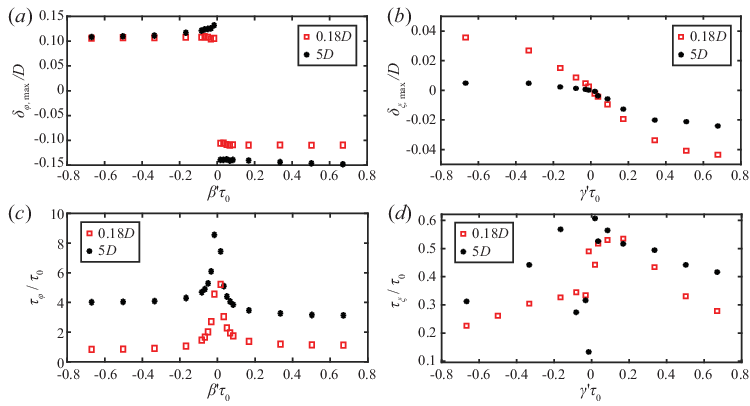}}
  \caption{(\textit{a}) Relationship between blade pitch rate and maximum wake expansion change at $0.18D$ and $5D$ downstream of the turbine. (\textit{b}) Relationship between yaw rate and maximum inverse wake deflection at $0.18D$ and $5D$ downstream. (\textit{c}) The effect of pitch rate on the wake expansion timescale at $0.18D$ and $5D$ downstream. (\textit{d}) The effect of yaw rate on the inverse wake deflection timescale at $0.18D$ and $5D$ downstream.}
\label{fig:5D}
\end{figure}

Quantitative investigation of dynamic wake modulation propagation downstream is also provided, \textcolor{blue}{though we note the limitations of this analysis. In the far wake, coherent structures have broken down and atmospheric turbulence is much more dominant than in the near wake \citep{Zhan2020}. Turbulence is not included in the current study to facilitate the observation of deterministic near wake modulation, as described in Section \ref{sec:les}. The downside of this simplification is that it limits the ability to draw realistic conclusions about far-wake propagation. Therefore, the following analysis adds primarily theoretical value for idealized conditions.} First, the wake expansion response to changes in blade pitch is calculated at $5D$ downstream of the rotor and compared to that observed at $0.18D$. \textcolor{blue}{The $5D$ distance is chosen because it is well within the far wake \citep[typically considered to begin at $2-4D$ downstream,][]{Gocmen2016}, but still within the higher-resolution portion of the simulation mesh (figure \ref{fig:mesh}\textit{a}). Furthermore, \citet{Santhanagopalan2018} found $5D$ to be the optimal spacing between the first and second row of turbines under most conditions.} To facilitate this far wake comparison, $\delta_{\varphi,\mathrm{max}}/D$, where $\delta_{\varphi}$ is the deviation of the wake diameter from $D$, is used to represent the amount of wake expansion rather than $\varphi_{\mathrm{max}}$. This metric is selected because its calculation does not depend on the downstream distance, unlike $\varphi$ (see figure \ref{fig:def}), and therefore demonstrates how the magnitude of a perturbation changes as it propagates downstream. As shown in figure \ref{fig:5D}(\textit{a}), the maximum wake expansion does not change significantly between $0.18D$ and $5D$. The maximum inverse wake deflection, on the other hand, decreases in magnitude as the perturbation is advected downstream (figure \ref{fig:5D}\textit{b}). The wake deflection metric used, $\delta_{\xi,\mathrm{max}}/D$, is defined using the spanwise deviation of the wake center (figure \ref{fig:def}), with the wake center location determined by a Gaussian fit to the streamwise wake velocity profile as described in section \ref{sec:fit}. Though figure \ref{fig:5D}(\textit{b}) shows a significant reduction in inverse wake deflection at $5D$, figure \ref{fig:down}(\textit{d}) clearly shows qualitatively that it does, in fact, persist into the far wake. This discrepancy between the qualitative and quantitative observations is attributed to the significant distortion of the wake cross-section in the far wake during periods of yaw misalignment \citep{howland2016,Bastankhah2016,Shapiro2018,Martinez-Tossas2019,Zong2020}.

\begin{figure}
  \centerline{\includegraphics{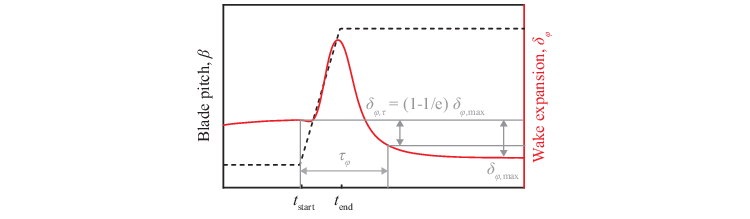}}
  \caption{Sample change in blade pitch angle at a constant rate (dashed black line) and the resulting wake expansion response at $5D$ downstream (solid red line), with the wake expansion timescale, $\tau_{\varphi}$, and maximum change in wake expansion, $\delta_{\varphi,\tau}$, defined in grey.}
\label{fig:expan5D}
\end{figure}

Wake expansion and deflection timescales are also calculated at $5D$ downstream and compared to those at $0.18D$, after accounting for the time it takes for the wake to be advected this distance by the ambient air. Figure \ref{fig:5D}(\textit{c}) shows that the wake expansion timescale increases significantly as the perturbation caused by changing blade pitch propagates downstream. This increase in timescale is caused by the phenomenon depicted in figure \ref{fig:down}(\textit{a}) and (\textit{b}). When the blade pitch increases, the maximum reduction in wake diameter occurs after the bulge caused by the velocity differences in the upwind and downwind portions of the wake passes. This bulge increases the time between the start of the pitch change maneuver and $\delta_{\varphi,\tau}$, the point used to determine $\tau_{\varphi}$ (figure \ref{fig:expan5D}). When the blade pitch decreases, the narrowing of the wake causes the same lag in wake expansion timescale. On the other hand, the deflection timescale at $5D$ downstream does not change significantly from that at $0.18D$ downstream (figure \ref{fig:5D}\textit{d}). Some increased scatter is observed, especially at the lower yaw rates, attributed to the relatively small inverse wake deflection at these yaw rates and the significant wake distortion discussed above, both of which lead to increased uncertainty in the definition of the wake deflection timescale.

\section{Conclusions and discussion}\label{sec:conc}
The current study uses LES to model transient behaviours in the wake of a wind turbine in response to changes in blade pitch, incoming wind direction, and yaw angle. Changes in blade pitch result in changes in wake expansion that are consistent with the results from the field study from \citet{Abraham2020}. The wake response to blade pitch changes displays hysteresis as a result of rotor and generator inertia, which causes a lag between changes in blade pitch and changes in tip speed ratio. The relationship between wake expansion and tip speed ratio exhibits the strongest dependence of the tested parameters, suggesting tip speed ratio is most responsible for changes in wake expansion. The lag between blade pitch and wake expansion changes is quantified for different pitch rates, showing that, though the maximum change in wake expansion depends on the magnitude of the pitch change, the time for the wake to stabilize varies significantly with pitch rate. Changes in wind direction are investigated to further elucidate the relationship between wake deflection and dynamic yaw misalignment observed in \citet{Abraham2020}, which was opposite of the steady yaw misalignment response reported in previous studies \citep[e.g.,][]{Jimenez2010,Bastankhah2016,Shapiro2018}. This behaviour is shown to depend strongly on the timescale of wind direction changes, with short-period changes (20 s) resulting in opposite wake deflection and long-period changes (50 s) demonstrating the steady wake deflection response. To separate the effects of turbine-modulated wake deflection from wake advection by the ambient flow, changes in rotor angle with a constant wind direction are investigated. The wake takes longer to reach the steady yaw response when the rotor angle is changed as a result of the removal of the wake advection effect. When the yaw rate is changed, a tradeoff is observed between inverse wake deflection time and magnitude. The mechanism behind the transient inverse wake deflection is elucidated by quantifying changes in streamwise vorticity in different parts of the wake. In the top-left quadrant, the vorticity increases during the yaw maneuver, while it decreases in the bottom-right quadrant, inducing a period of positive spanwise velocity. These behaviours are caused by changes in the relative velocities of each part of the rotor and the resulting change in projection angle of the blade tip vortices. Finally, these near-wake behaviours are shown to propagate downstream to the far wake. The wake expansion timescale increases as the perturbation induced by the blade pitch change is advected downstream, while the wake deflection timescale remains largely unchanged.

The results of the current study provide insight into the wake behaviour of utility-scale wind turbines operating under atmospheric conditions, enabling more accurate modelling and prediction of the real-world wake. They explain phenomena previously observed in high-resolution field scale wake data. Furthermore, they have significant implications for recently proposed dynamic flow control strategies. One such strategy is the regulation of the thrust coefficient of each turbine in a wind farm, e.g., by modifying the blade pitch, to optimize the overall farm power generation \citep{Goit2015,Munters2017,Yilmaz2018}. This method can also be applied for power tracking, allowing wind farms to provide additional grid services \citep{Shapiro2017}. Another strategy for wind farm optimization is adaptive yawing to dynamically deflect wind turbine wakes away from downstream turbines \citep{Gebraad2015,Raach2017,Raach2018}. \citet{Munters2018} proposed combining these two strategies to fully optimize wind farm power output, and \citet{Kanev2018} included fatigue loading considerations to maximize overall wind farm lifetime. However, the aforementioned studies did not provide detailed analyses of the timescales of the proposed dynamic wake changes. In order to effectively implement these novel wake control strategies by dynamically varying parameters such as blade pitch and rotor yaw, the transient wake response must be included in the models used to develop such algorithms. The current study shows that the wake consistently takes time on the order of \textcolor{blue}{$D/U_\infty$ ($\sim$10 s for the setup presented here)} to reach a steady response. Therefore, wake changes with timescales shorter than this cannot be implemented effectively. This limitation is determined by the physical timescales of the vorticity in the wake and the inertia of the rotor.

Dynamic wake modulation is expected to enhance wake recovery downstream due to the fluctuations of the near-wake boundary by an average 11\% increase in energy flux into the wake, as discussed in \citet{Abraham2020} and \citet{Abraham2020_1}. Though turbulence still dominates wake recovery, dynamic wake modulation provides a significant contribution. \textcolor{blue}{\citet{Munters2018b},\citet{Yilmaz2018}, and \citet{Frederik2020} exploited dynamic wake behaviors to further enhance wake breakdown and accelerate recovery. \citet{Brown2021} also showed that dynamically varying the blade pitch and rotor speed excites unstable blade tip vortex modes, accelerating breakdown.} The results of the current study additionally suggest that these dynamic wake modulations persist into the far wake where they can further impact on wake mixing. Furthermore, when the rotor misalignment angle changes, the resulting asymmetry of the blade tip vortices is expected to accelerate tip vortex breakdown. Tip vortex breakdown is typically induced by the development of instabilities in the vortex helix, particularly by the mutual inductance (vortex pairing) instability \citep{Ivanell2010,Felli2011,Sorensen2011,Sarmast2014,Lignarolo2015}. The fundamental study of \citet{Ortega2003} and the modal decomposition investigation of \citet{Sarmast2014} show that asymmetric vortex configurations destabilize faster than symmetric configurations. As tip vortex breakdown is the first step of wake recovery, the tip vortex asymmetry due to rotor misalignment will contribute to the enhanced wake recovery caused by dynamic wake modulation. 

The asymmetric wake response to changes in rotor misalignment is also expected to contribute to wake meandering. Currently, two mechanisms have been proposed in the literature to understand wake meandering: the dynamic wake meandering (DWM) model which considers the wake to be a passive tracer advected by large-scale atmospheric motions, and the idea that the rotor acts like a bluff body shedding vortices \citep{Yang2019}. Both of these frameworks treat the turbine as a passive structure, but our results show that the wake deflection is actively modulated by the turbine. In the field, the yaw misalignment between the rotor and wind is constantly changing due to the constant changes in wind direction, which leads to continual changes in wake deflection. These turbine-modulated wake deflections will be superimposed on any meandering caused by advection or bluff body vortex shedding.

\textcolor{blue}{The findings of the current study also have implications for turbines undergoing tilt angle changes. These tilt changes can be induced either by the rocking motion of ocean waves for offshore floating turbines \citep{Rockel2014} or by active wake control \citep{Annoni2017,Su2020}. Previous studies have shown that the magnitude of wake deflection in the vertical direction during rotor tilt is comparable to spanwise wake deflection during yaw \citep{Bossuyt2021}, and that the counter-rotating vortex mechanism responsible for inducing steady-state spanwise wake deflection is also responsible for tilt-induced vertical wake deflection \citep{Su2020}. These similarities in steady-state behaviors suggest analogous dynamic wake behaviors would result from dynamic tilt changes. For floating offshore turbines, all tilt changes are dynamic, as wave-induced motions are inherently oscillatory. Alternatively, dynamic tilt-based control algorithms could be proposed with the addition of the necessary mechanical components. Future studies may investigate whether the potential benefits of such a control method outweigh the added cost of implementation.}

Finally, we acknowledge that the current study investigates the wake of a single example turbine. We expect the qualitative trends described here to hold true regardless of turbine design, but we caution the reader that the quantitative measures may not translate directly to other turbines. Future work can investigate the potential impact of turbine size, geometry, rotation direction, and siting on the observed phenomena. Additionally, the wake behavior was studied under uniform inflow in order to isolate the effects of each parameter changed individually. However, the effect of such atmospheric phenomena as wind shear and veer can be investigated in future studies. More significantly, turbulence has not been included in the current study. Though the near wake is dominated by the coherent structures shed from the turbine, atmospheric turbulence dominates far-wake behaviour in the field \citep{Zhan2020}. Therefore, the direct applicability of the downstream dynamic wake modulation propagation analysis to wind turbines operating under real-world conditions is limited. \textcolor{blue}{As discussed in Section \ref{sec:down}}, this far-wake analysis adds primarily theoretical value for idealized conditions.

\section*{Acknowledgements}
The research was performed using computational resources sponsored by the Department of Energy’s Office of Energy Efficiency and Renewable Energy and located at the National Renewable Energy Laboratory. Funding was provided by the National Science Foundation INTERN funding opportunity as a supplement to NSF CAREER award NSF-CBET-1454259. The authors also thank N. Hamilton, S. Ananthan, G. Vijayakumar, and M. Sprague from NREL for fruitful discussions and assistance with the simulations.

\section*{Declaration of interests}
The authors report no conflict of interest.

\appendix
\section{}\label{appA}
\begin{figure}
  \centerline{\includegraphics{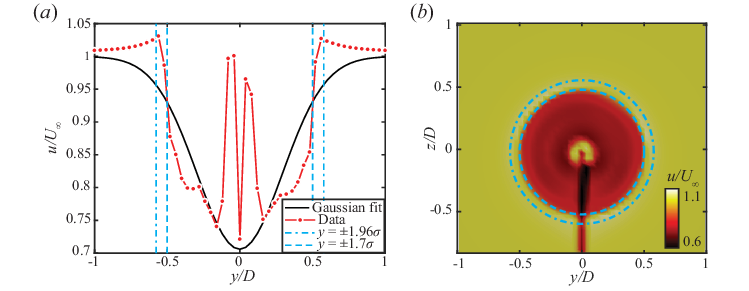}}
  \caption{(\textit{a}) Cross-section at hub height of the wake velocity field for a sample simulation time step compared to the corresponding Gaussian fit. Note that the sharp velocity peaks in the middle are characteristic of flow acceleration around the turbine nacelle. The width of the 95\% ($y=\pm1.96\sigma$) and 91\% ($y=\pm1.7\sigma$) confidence intervals are also indicated as vertical lines. (\textit{b}) Wake velocity field with 95\% (dot-dashed circle) and 91\% (dashed circle) confidence intervals superimposed.}
\label{fig:app}
\end{figure}

The need for the selection of the 91\% confidence interval to define the wake width in the near wake rather than the 95\% interval typically used for the far wake becomes clear by looking closely at the Gaussian fit for a sample timestep. In the near wake, the velocity deficit drops more sharply at the wake edges than the Gaussian function. A cross-section of the wake velocity profile and Gaussian fit at hub height is shown in figure \ref{fig:app}(\textit{a}), along with lines indicating the location of the 95\% confidence interval ($1.96\sigma$) and the 91\% confidence interval ($1.7\sigma$). The 95\% interval intersects the velocity profile outside of the region of the velocity deficit that defines the wake. On the other hand, the 91\% interval intersects the velocity profile at the point of maximum velocity gradient, which demarcates the wake boundary. The specific value of 91\% was selected by finding the point of maximum velocity gradient at multiple timesteps. The need for the shift in definition of the wake boundary is further clarified in figure \ref{fig:app}(\textit{b}), which shows the near wake velocity cross-section with both confidence intervals superimposed. The outer is the 95\% interval, which clearly overshoots the wake boundary, while the inner circle (91\% interval) captures the wake edge much more accurately.

\section{}\label{appB}
\textcolor{blue}{Figure \ref{fig:stills} shows streamwise vorticity in the near wake of the turbine during key points of the rotor yawing maneuver. The beginning of the yawing maneuver and the resulting increase in streamwise vorticity in the top-left quadrant of the wake are shown in the top two frames. The decrease in streamwise vorticity in the top-left quadrant of the wake due to the steady wake response and the end of the yawing maneuver are shown in the bottom two frames.}

\begin{figure}
  \centerline{\includegraphics{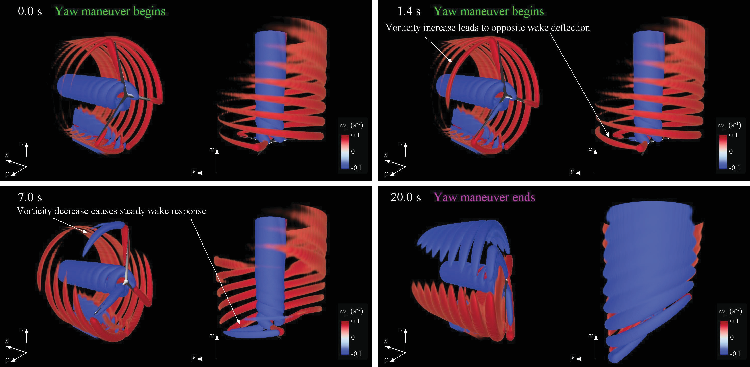}}
  \caption{\textcolor{blue}{Stills of key points in supplementary movie 3 showing isocontours of streamwise vorticity in the wake during a rotor yawing maneuver. The values of the isocontours are $\omega_x=0.1$ s\textsuperscript{-1} and $\omega_x=-0.1$ s\textsuperscript{-1} for the positive (red) and negative (blue) isocontours, respectively.}}
\label{fig:stills}
\end{figure}

\bibliographystyle{jfm}
\bibliography{JFMR2020}

\end{document}